\documentclass[pre,superscriptaddress,showpacs,amsmath,twocolumn]{revtex4}  %% REVTeX 4.0
\newcommand{\beq}{\begin{equation}}
\newcommand{\enq}{\end{equation}}

\usepackage{graphicx}% Include figure files

\begin{document}

\title{Beyond the random phase approximation: Stimulated Brillouin backscatter  for finite laser coherence times}

\author{Alexander O. Korotkevich}
\affiliation{Department of Mathematics and Statistics, University of New Mexico, Albuquerque, NM 87131, USA}
\affiliation{Landau Institute for Theoretical Physics, 2 Kosygin Str., Moscow, 119334, Russia}
\author{Pavel M. Lushnikov}
\email{plushnik@math.unm.edu}
\affiliation{Department of Mathematics and Statistics, University of New Mexico, Albuquerque, NM 87131, USA}
\author{Harvey A. Rose}
\affiliation{New Mexico Consortium, Los Alamos, New Mexico 87544, USA}
%\affiliation{Theoretical Division, Los Alamos National Laboratory,  MS-B213, Los Alamos, New Mexico, 87545}

\date{%Printed
\today
%November 19, 2003
}

\begin{abstract}
We develop a statistical theory of  stimulated Brillouin backscatter (BSBS)  of a spatially and temporally partially incoherent laser beam for laser fusion relevant plasma. We find
a new collective regime of BSBS (CBSBS) with intensity threshold controlled by diffraction, an insensitive function  of the laser coherence time, $T_c$, once light travel time during
$T_c$ exceeds a laser speckle length.
The BSBS spatial gain rate is approximately the sum of that due to CBSBS, and a part which is independent of diffraction and varies linearly with $T_c$.
We find that the bandwidth of  KrF-laser-based fusion systems would be large enough to allow additional suppression of BSBS.
\end{abstract}


\maketitle

\section{Introduction}

Inertial confinement fusion (ICF) experiments require propagation of intense laser light through underdense plasma subject to laser-plasma instabilities which can be deleterious for
achievement of thermonuclear target ignition because they can cause the loss of target symmetry, energy and hot electron production \cite{Lindl2004}.
Among laser-plasma instabilities,  backward stimulated Brillouin scatter (BSBS) has long  been considered a serious danger because the damping threshold of
BSBS of coherent laser beams is  typically several order of magnitude less then the required laser intensity $\sim  10^{15}\mbox{W}/\mbox{cm}^2$ for ICF. BSBS
may result in laser energy retracing its path to the laser optical
system, possibly damaging laser components \cite{Lindl2004,MeezanEtAlPhysPlasm2010}.
Recent experiments for a first time achieved conditions of fusion plasma and indeed demonstrated that large levels of BSBS (up to tens percent of reflectivity) are possible \cite{FroulaPRL2007}.

Theory of laser-plasma interaction  instabilities is well developed for coherent laser beam \cite{Kruer1990}.
%However,
%ICF laser beams are not coherent and classic theory is not applicable\cite{Kruer1990,Lindl2004,MeezanEtAlPhysPlasm2010}.
However, ICF laser beams are not coherent because temporal and spatial beam smoothing techniques are currently used to produce laser beams with
short enough correlation time, $T_c,$  and lengths to suppress self-focusing \cite{Kruer1990,Lindl2004,MeezanEtAlPhysPlasm2010}.
The laser intensity forms a speckle field - a random in space distribution of intensity with transverse correlation length
$l_c\simeq F\lambda_0$ and longitudinal correlation length (speckle length) $L_{speckle}\simeq 7F^2\lambda_0$, where $F$ is the optic $f$-number and $\lambda_0=2\pi/k_0$ is the wavelength
(see e.g. \cite{RosePhysPlasm1995,GarnierPhysPlasm1999}).
There is a long history of study of amplification in random media (see e.g \cite{vedenov1964,PesmeBerger1994} and references there in).  For small laser beam
  correlation time $T_c$, the spatial instability increment
 is given by a Random Phase Approximation (RPA).
 Beam smoothing for ICF typically has $T_c$ much above the regime of RPA applicability. There are few examples in which implications of
  laser beam spatial and temporal incoherence have been analyzed for such larger $T_c$. One exception is forward stimulated Brillouin scattering (FSBS).  %FSBS for both a strictly coherent
 % laser beam and RPA limit is a classic linear theory.
 We have  obtained in Refs. \cite{LushnikovRosePRL2004,LushnikovRosePlasmPhysContrFusion2006} the FSBS dispersion relation for laser beam
  which has the correlation time $T_c$  too large for RPA relevance, but still small enough to suppress single laser speckle instabilities \cite{RoseDuBois1994}.
  We verified our theory of this ``collective" FSBS instability regime with 3D simulations. Similar simulation results had been previously observed \cite{SchmittAfeyan1998}.

 This naturally leads one to consider the possibility of a collective regime for BSBS  (CBSBS). We present 2D and 3D simulation results as evidence for such a regime, and find
   agreement with a simple theory  that above CBSBS threshold, the spatial increment for
  backscatter amplitude $\kappa_i$, is well approximated by the sum of two contributions.   The first
contribution is RPA-like  $\propto T_c$ without intensity threshold (we neglect light wave damping). The second contribution
 has a threshold in laser intensity.
That threshold is in  parameter range of ICF hohlraum plasmas such as
at  the National Ignition Facility (NIF) \cite{Lindl2004} and the Omega
laser facility (OMEGA) \cite{NiemannPRL2008} experiments. The existence of threshold was first predicted in Ref. \cite{LushnikovRoseArxiv2007}
in the limit $cT_c\gg L_{speckle},$ where $c$ is the speed of flight \cite{RemarkReplete}.
The second contribution is collective-like because
it neglects speckle contributions and is only weakly dependent on $T_c$.
CBSBS threshold is applicable for strong and weak acoustic damping coefficient $\nu_{ia}$.
The theory also demonstrates
 a good quantitative prediction of the instability increment for small $\nu_{ia}\sim 0.01$
  which is relevant for gold
plasma near the wall of hohlraum in  NIF and OMEGA experiments \cite{Lindl2004,NiemannPRL2008}.

The paper is organized as follows. In Section \ref{sec:Basicequations} we introduce the basic equations of BSBS for laser-plasma interaction and the stochastic boundary conditions
which correspond to the partial incoherence of laser beam.  In Section \ref{sec:Linearinstability} we analyze the
linearized BSBS equations and find the dispersion relations.
In Section \ref{sec:ConvectiveAbsolute} the convective versus absolute instabilities are analyzed from the dispersion relations.
Section \ref{sec:Numericalsimulations} describes the details of the performed stochastic simulations of the full linearized equations.
In section  \ref{sec:Applicability}
the conditions of applicability of the dispersion relation are discussed as well as the  estimates for  typical ICF experimental conditions are given.
In Section \ref{sec:Conclusion} the main results of the paper are
discussed.

\section{Basic equations}
\label{sec:Basicequations}

Assume that laser beam propagates in plasma with frequency
$\omega_0$  along $z$. The  electric field
$\cal E$ is given by
\begin{eqnarray}\label{EBdef}
{\cal E}=(1/2)e^{-i\omega_0 t}\Big [E e^{ik_0 z}+Be^{-ik_0
z-i\Delta\omega t}\Big ]+c.c.,
\end{eqnarray}
where $E({\bf r}, z,t)$ is the envelope of laser beam and $B({\bf
r}, z,t)$ is the envelope of backscattered wave,  ${\bf r}=(x,y)$,
and c.c. means complex conjugated terms. Frequency shift $\Delta
\omega=-2k_0c_s$ is determined by coupling of $E$ and $B$ through ion-acoustic wave of phase speed
$c_s$ and wavevector $2k_0$ with  plasma density fluctuation
$\delta n_e$ given by $\frac{\delta n_e}{n_e}=\frac{1}{2}\sigma
e^{2ik_0z+i\Delta\omega t}+c.c.,$ where $\sigma({\bf r}, z,t)$ is
the slow envelope (slow provided $\Delta\omega T_c\gg 1$)  and $n_e$ is the average electron density, assumed
small compared to the critical electron density $n_c$. We consider a slab model of plasma (plasma parameters are
uniform). The coupling of $E$
and $B$ to plasma density fluctuations gives
\begin{eqnarray}\label{EBeq1}
R_{EE}^{-1}E \equiv  \left [ i\Big (c^{-1}{\partial_t}+{\partial_z}\Big )+\frac{1}{2k_0}\nabla^2
  \right ]E=\frac{k_0}{4}\frac{n_e}{n_c}\sigma B, \\
R_{BB}^{-1}B  % \qquad  \qquad  \qquad  \qquad  \qquad  \qquad  \qquad  \qquad  \qquad  \qquad  \nonumber \\
\equiv  \left [ i\Big (c^{-1}{\partial_t}-{\partial_z}\Big )+\frac{1}{2k_0}\nabla^2
  \right ]B=\frac{k_0}{4}\frac{n_e}{n_c}\sigma^* E, \label{EBeq2}
\end{eqnarray}
 $\nabla=({\partial_x},{ \partial_y})$, and $\sigma$ is described by the acoustic wave equation coupled to
the pondermotive force $\propto {\cal E}^2$ which results in the
envelope equation
\begin{eqnarray}\label{sigma1}
R_{\sigma\sigma}^{-1}\sigma^* \equiv   [ i ({c_s^{-1}}{\partial_t}+2\nu_{ia} k_0+{\partial_z} )-(4k_0)^{-1}\nabla^2
   ]\sigma^* \nonumber \\
   =-2k_0  E^*B.
\end{eqnarray}
%
%The response of the slowly varying part of $\delta n_e$ to the slowly varying %part of the ponderomotive force ($\propto|E|^2 + |B|^2$), which is responsible %for self-focusing, is neglected.
Here we neglected terms $\propto |E|^2, \ |B|^2$ in the right-hand side  (r.h.s.) which are responsible for self-focusing effects,
$\nu_L$ is the Landau damping of ion-acoustic wave and
$\nu_{ia}=\nu_L/2k_0c_s$ is the scaled acoustic Landau damping coefficient.  $E$ and $B$ are in thermal units (see e.g. \cite{LushnikovRosePRL2004}) defined so
that if we add self-focusing term $I=|E|^2$ in r.h.s. of
Eq. (\ref{sigma1}) then in equilibrium, with uniform $E$, the
standard $\delta n_e/n_e=\exp (-I)-1$ is recovered.

We use a simple model of induced spacial incoherence beam smoothing \cite{LehmbergObenschain1983} which defines stochastic boundary conditions at $z=0$ for
spatial Fourier transform (over ${\bf r}$) components $ \hat E({\bf k})$, of laser beam amplitude
\cite{LushnikovRosePRL2004}:
\begin{eqnarray}\label{phik}
&\hat E({\bf k },z=0,t)= |E_{\bf k}|\exp [ i\phi_{\bf
k}(t) ], \nonumber \\
&  \langle \exp i [\phi_{\bf
k}(t)-\phi_{{\bf k}'}(t') ] \rangle =\delta_{{\bf k
k}'}\exp (-|t-t'|/T_c),
\end{eqnarray}
where
\begin{eqnarray}\label{phiktophat}
 |E_{\bf k}|=const, \ k<k_m; \  E_{\bf k}=0, \ k>k_m,
\end{eqnarray}
is chosen as the idealized ``top hat" model of NIF optics \cite{polarization}. Here $\langle \ldots \rangle$ means the averaging over the ensemble of stochastic realizations of boundary conditions,
$k_m\simeq k_0/(2F)$ is the top hat width
and the average intensity, $   \langle I
\rangle \equiv  \langle |E|^2 \rangle =I$ determines the constant.

\section{Linearized equations and dispersion relations}
\label{sec:Linearinstability}

In linear approximation, assuming $|B|\ll |E|$ so that only the laser beam is BSBS unstable, we neglect  right hand side (r.h.s.) of Eq. (\ref{EBeq1}).
 The
resulting linear equation with  boundary condition (\ref{phik})  has the exact solution as
 decomposition of $E$ into
 Fourier series,
\begin{eqnarray}\label{Elin}
E({\bf r},z,t)=\sum_j E_{{\bf k}_j}, \nonumber \\
%\end{eqnarray}
%
%with
E_{{\bf k}_j} =|E_{{\bf k}_j}| \exp\big [ i(\phi_{{\bf k}_j}(t-z/c)+{\bf k}_j\cdot {\bf r}-{\bf k}_j^2z/2k_0)\big ].
\end{eqnarray}

\begin{figure}%[htbp]
\begin{center}
\includegraphics[width = 3.45 in]{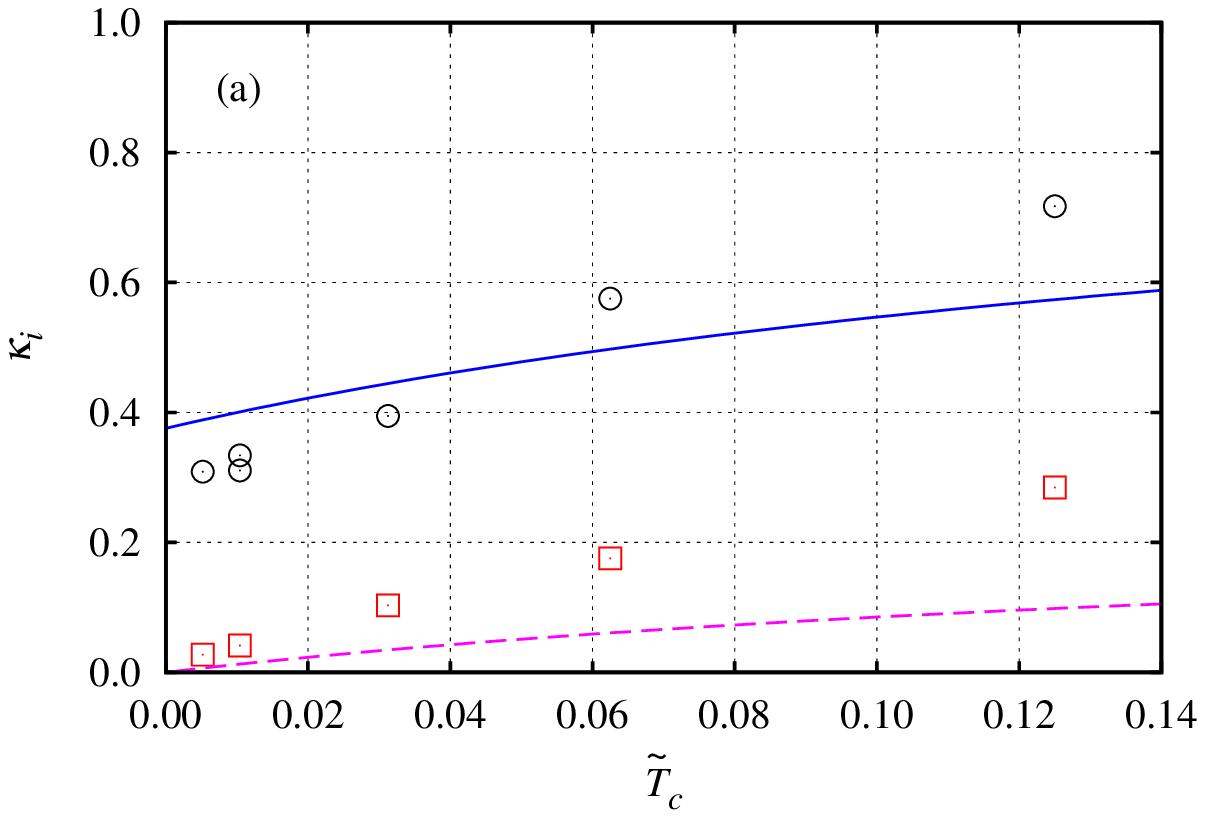} \\
\includegraphics[width = 3.45 in]{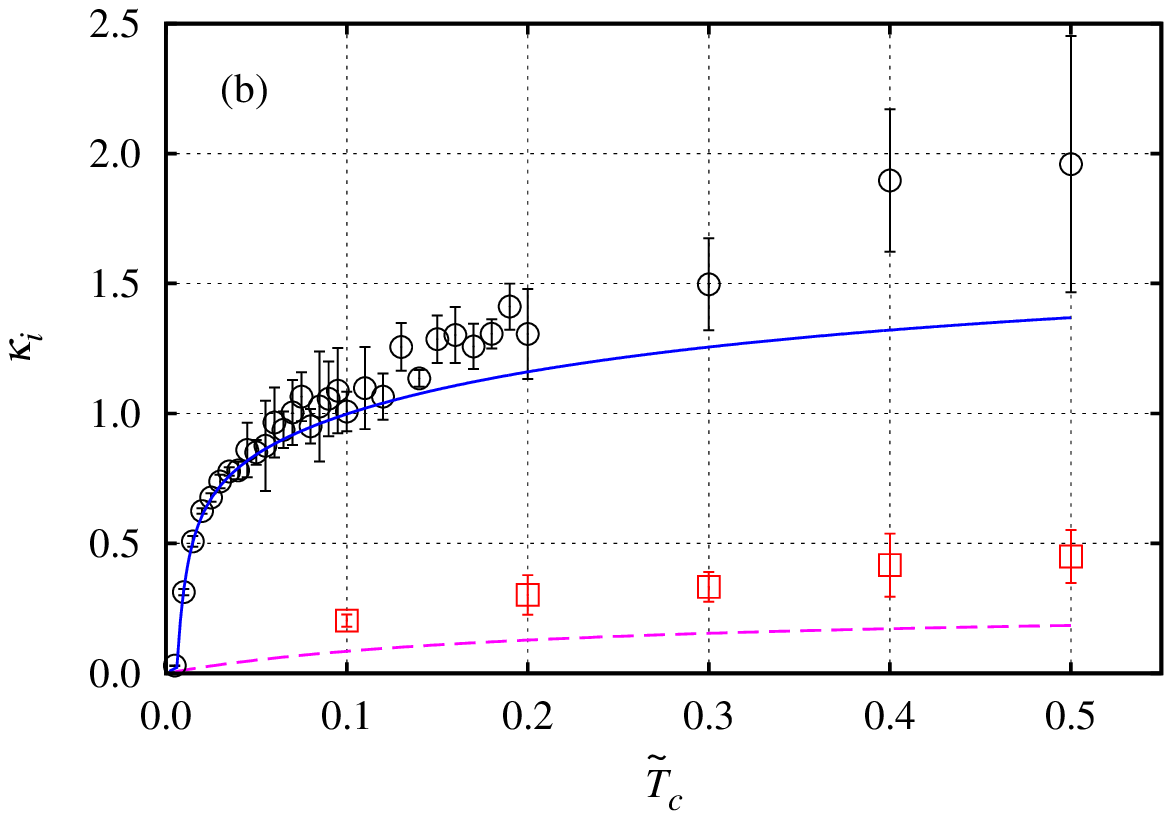}
\end{center}
\caption{Spatial increment $\kappa_i$ of CBSBS obtained from stochastic  simulations of (\ref{EBeq2})-(\ref{Elin})
compared with the sum of increments
$\kappa_B+\kappa_\sigma$ (obtained by solving (\ref{dispk0contTc}) and (\ref{dispk0contTcB})).
The scaled damping rate  $\mu=5.12$ is used (e.g. it corresponds to  $\nu_{ia}=0.01, \ F=8$). (a) $3D$ simulations with
$c_s/c=0$, $\tilde I=2$ (circles) and $\tilde I=1$ (squares). The scaled dimensionless laser intensity $\tilde I$, $\mu$ and the scaled correlation time $\tilde T_c$ are defined in (\ref{Itildedef}).
Solid and dashed lines show $\kappa_B+\kappa_\sigma$ for  $\tilde I=2$ and $\tilde I=1$, respectively. If $\kappa_\sigma<0$ then
$\kappa_B+\kappa_\sigma$ is replaced by $\kappa_B$.
(b) $2D$ simulations   with the modified boundary conditions, $c/c_c=500$,   $\tilde I=3$ (circles) and $\tilde I=1$ (squares). Error bars  are also shown.
Solid and dashed lines show $\kappa_B+\kappa_\sigma$ for  $\tilde I=3$ and $\tilde I=1$, respectively, for both (a) and (b). The details of simulation method are provided in Section \ref{sec:Numericalsimulations}.
}
\vspace{0.1cm}
\label{fig:figkappa}
\end{figure}

Figures  \ref{fig:figkappa} show the  increment $\kappa_i$ % (in the units $k_m^2/k_0$)
of the spatial growth of backscattered light intensity $\langle |B|^2\rangle \propto e^{-2\kappa_i z}$ as a
 function of the rescaled correlation time   $\tilde T_c$
  obtained from the numerical solution of the stochastic linear equations (\ref{EBeq2})-(\ref{Elin})
 (details of numerical simulations are provided in Section \ref{sec:Numericalsimulations}), the scaled damping rate $\mu$  and the scaled %dimensionless
laser intensity $\tilde I$. These scaling quantities are  defined as
\begin{equation}\label{Itildedef}
 \tilde T_c\equiv T_c k_0c_s /4F^2, \ \mu\equiv 2\nu_{ia} k_0^2/k_m^2, \ \tilde I\equiv\frac{4F^2}{\nu_{ia}}\frac{n_e}{n_c} I.
 \end{equation}
Here $\tilde T_c$ has the meaning  of the correlation time $T_c$ in units of the acoustic transit time along speckle.  (Note that definition of $\tilde T_c$
  is different by a factor $1/2F$ from the definition used for FSBS
 \cite{LushnikovRosePRL2004,LushnikovRosePlasmPhysContrFusion2006}, where units of the transverse acoustic transit time through speckle were used.)
We use dimensionless units with  $k_0/k_m^2=4F^2/k_0$ as the unit in $z$ direction,
and $k_0/k_m^2 c_s$ is the time unit.
$\langle \ldots \rangle$ means averaging over the statistics of laser beam fluctuations  (\ref{phik}).
$\mu$ is the damping rate in units of the inverse acoustic propagation time along a speckle. (See also Figure \ref{fig:kappaI} below for illustration of intensity
normalization in comparison with physical units.)

\begin{figure}[b]
\begin{center}
\includegraphics[width = 3.45 in]{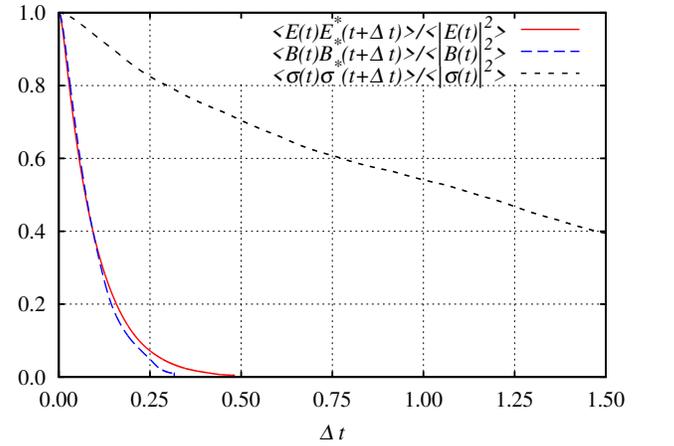}
%\vspace{0.5cm}
\end{center}
\caption{
Normalized autocorrelation functions vs. a dimensionless time shift $\Delta  t$ for $E, \ B$ and $\sigma$:  $\langle E({\bf r},z,t)E^*({\bf r},z,t+\Delta  t)\rangle $,
$\langle B({\bf r},z,t)B^*({\bf r},z,t+ \Delta  t)\rangle $ and $\langle \sigma({\bf r},z,t)\sigma^*({\bf r},z,t+\Delta  t)\rangle $ % in time for
with $\tilde I=3,$ $\tilde T_c=0.1$ and $\mu=5.12$ from stochastic  simulations of (\ref{EBeq2})-(\ref{Elin}).
It is seen that $B$ is correlated at  the same time $\tilde T_c$ as $E$ while $\rho$ is correlated at much larger times.
}
\label{fig:figcorrelation}%\label{fig:fig1}
\end{figure}

We  relate $\kappa_i$ to the instability increments for   $\langle B \rangle$ and $\langle \sigma^* \rangle$ (we designate them $\kappa_B$ and $\kappa_\sigma$, respectively).
In general, growth rates of mean amplitudes give a lower bound to $\kappa_i$. However, according to Figure \ref{fig:figcorrelation}, $\sigma$ is almost coherent on a time scale $T_c$
justifying the use of mean values of amplitudes.
%We now compare these numerical results with the simple theory based on the separation of the contributions to the spatial increment $\kappa_i$ from the

First we look for the expression for $\kappa_\sigma$.  Eq. (\ref{EBeq2}) is linear in $B$ and $E$  implying that
$B$ can be decomposed into $B=\sum_j B_{{\bf k}_j}$
with
\begin{eqnarray}\label{Bjeq}
 R_{BB}^{-1}B _{{\bf k}_j}=\frac{k_0}{4}\frac{n_e}{n_c}\sigma^* E_{{\bf k}_j}.
  \end{eqnarray}
Approximating r.h.s. of (\ref{sigma1}) as $E^*B\simeq \sum_j
E_{{\bf k}_j}^*
B_{{\bf k}_j}$ gives
\begin{eqnarray}\label{sigma2}
 % \left
 % [ i (\frac{1}{c_s}{\partial_t}+2\nu_{ia} k_0+{\partial_z})-\frac{1}{4k_0}\nabla^2
  %\right
 % ]
 R_{\sigma\sigma}^{-1}\sigma^*%\nonumber \\
  =-2k_0  \sum_jE_{{\bf k}_j}^*B_{{\bf k}_j},
\end{eqnarray}
which means that we neglect off-diagonal terms $E_{{\bf k}_j}^*
B_{{{\bf k}_j}'}, \quad j\neq j'.$ Since speckles of
laser field arise from interference of different Fourier modes,
$j\neq j',$ we associate the off-diagonal terms with speckle
contribution to BSBS
\cite{RoseDuBois1993,RosePhysPlasm1995,RoseMounaixPhysPlasm2011}.
Neglecting   off-diagonal terms requires that during time $T_c$ light travels much further than a speckle length, $L_{speckle}\ll cT_c$
  and that $T_c \ll t_{sat}$, where $t_{sat}$ is the characteristic time scale at which
BSBS convective gain saturates at each speckle \cite{MounaixPRL2000}.

Eqs. (\ref{Bjeq}) and (\ref{sigma2}) result in
\begin{eqnarray}\label{sigmastar}
R_{\sigma\sigma}^{-1}\langle \sigma^*\rangle =-(k_0^2/2)(n_e/n_c)\langle \sum_j E_{{\bf k}_j}^*   R_{BB}\sigma^* E_{{\bf k}_j}  \rangle
\end{eqnarray}
with the Fourier transformed  $R_{BB}$ given by
\begin{eqnarray}\label{Rbb0}
\hat R_{BB}({\bf k},z,t) =-ic\delta(z+ct)\exp{[i\frac{k^2}{2k_0}z]\Theta(-z)},
\end{eqnarray}
where $\Theta(z)$ is the Heaviside step function.

We assume that $\sigma^*$ is slow in comparison with $ E_{{\bf k}_j}$ (consistent with Figure \ref{fig:figcorrelation})
which allows to approximate fluctuating terms in r.h.s. of \eqref{sigmastar} as  $\langle E_{{\bf k}_j}^*   \sigma^* E_{{\bf k}_j}\rangle\simeq \langle \sigma^* \rangle \langle E_{{\bf k}_j}^* E_{{\bf k}_j}\rangle$ which has the same form as the Bourret
approximation \cite{PesmeBerger1994} and provides the closed expression for $\langle \sigma^* \rangle$ as follows
\begin{eqnarray}\label{sigmastaraveraged}
R_{\sigma\sigma}^{-1}\langle \sigma^*({\bf r},z,t)\rangle
=-(k_0^2/2)(n_e/n_c)\int\int\int d{\bf r}'dz'dt'
\nonumber \\
\times R_{BB}({\bf r}-{\bf r}',z-z',t-t') C^*({\bf r}-{\bf r}',z-z',t-t')
\nonumber \\
\times\langle \sigma^* ({\bf r}',z',t') \rangle,
\end{eqnarray}
where the kernel of the response function $R_{BB}({\bf x},z,t)$ is the inverse Fourier transform of \eqref{Rbb0} and the laser beam correlation function $C$ is given by
\begin{eqnarray}
%
%\begin{align}
%\begin{split}
\label{cdef}
C({\bf r}-{\bf r}',z-z',t-t')\equiv\langle E({\bf r},z,t)E^*({\bf r}',z',t')\rangle
\nonumber \\
=\sum_j|E_{{\bf k}_j}|^2\exp\Big [i{\bf k}_j\cdot({\bf r}-{\bf r}')-i\frac{{\bf k}_j^2}{2k_0}(z-z')
\nonumber \\
-\big |t-t'-(z-z')/c\big |/T_c\Big ]
%\end{split}
%\label{cdef}
%\end{align}
\end{eqnarray}
for the top hat model \eqref{phik},\eqref{phiktophat} and \eqref{Elin}.

 We look for solution of \eqref{sigmastaraveraged} in exponential form
$\langle\sigma^*\rangle \propto e^{i(\kappa z+{\bf k}\cdot {\bf r}-\omega t)}$, then the exponential time dependence
of (\ref{cdef}) allows to carry all integrations in \eqref{Rbb0} and \eqref{sigmastaraveraged} explicitly to arrive at the following
dispersion relation in dimensionless units
\begin{eqnarray}\label{dispk1}
-i\omega+\mu+i\kappa-(i/4)k^2 \qquad  \qquad  \qquad  \qquad  \qquad  \qquad   \nonumber \\
=8iF^4\frac{n_e}{n_c}\sum\limits_{j=1}^N \frac{|E_{{\bf k}_j}|^2}{\omega\frac{c_s}{c}+\kappa-k_j^2-\frac{k^2}{2}-{\bf k}_j\cdot {\bf k}+2i\frac{c_s}{c}\frac{1}{\tilde T_c}},
\end{eqnarray}
%
%where $\mu\equiv 2\nu_{ia} k_0^2/k_m^2,$ $1/k_m$ is the transverse unit of length, $k_0/k_m^2$ is the unit in $z$ direction and $k_0/k_m^2 c_s$ is the time unit.
where % $1/k_m$ is the transverse unit of length and
vectors ${\bf k}_j$ span  top hat (\ref{phik}),\eqref{phiktophat}, and  $I=\sum_j|E_{{\bf k}_j}|^2$.

In the continuous limit $N\to \infty$, sum in (\ref{dispk1}) is replaced by integral, giving for the most unstable mode ${\bf k}=0$:
\begin{eqnarray}\label{dispk0contTc}
\Delta_\sigma(\omega,\kappa)&&=
-i\omega+\mu+i\kappa\nonumber \\
&&
+i\frac{\mu}{4}\tilde I\ln\frac{1-\kappa-\omega\frac{c_s}{c}-2i\frac{c_s}{c}\frac{1}{\tilde T_c}}{-\kappa-\omega\frac{c_s}{c}-2i\frac{c_s}{c}\frac{1}{\tilde T_c}}%\nonumber \\
=0,
\end{eqnarray}
which supports the convective instability with the increment $\kappa_\sigma\equiv Im(\kappa)>0$ only for $\tilde I> \tilde I_{convthresh}$, where $\tilde I_{convthresh}$
is the convective CBSBS threshold given by
\begin{eqnarray}\label{I0convthresh}
\tilde I_{convthresh}=\frac{4}{\pi}\Big ( 1-\frac{8c_s}{\pi c\tilde T_c}\Big )^{-1}.
\end{eqnarray}
In the limit $c/c_s \to \infty$,  the increment $\kappa_\sigma$ is independent of $\tilde T_c$  which suggests that we refer to it as the collective instability branch.
For finite but small $c_s/c\ll 1$ and $\tilde I>\tilde I_{convthresh}$  there is sharp transition of $\kappa_\sigma$ as a function of $\tilde T_c$ from 0 for $\tilde T_c=0$ to
 $\tilde T_c$-independent value of $\kappa_\sigma$. That value can be obtained analytically from (\ref{dispk0contTc})
 for $I$ just above the threshold  as follows: $\kappa_\sigma=\mu(\pi/4-2\tilde T_c^{-1}c_s/c)(\tilde I-\tilde I_{convthresh})/(\mu \tilde I-1)$ \cite{absolutecomment}.

The increment $\kappa_B$ is obtained in a similar way by statistical averaging of equation (\ref{EBeq2}) for $\langle B\rangle $ which gives
\begin{eqnarray}\label{Baverage}
R_{BB}^{-1}\langle B\rangle =-(k_0^2/2)(n_e/n_c)\langle E   R_{\sigma\sigma}E^*B\rangle
\end{eqnarray}
with  the Fourier transformed response function
\begin{align} \label{Rsigmasigma0}
\hat R_{\sigma\sigma}({\bf k},z,t) =-ic_{s}\delta(z-c_st)
\exp{[(i\frac{k^2}{4k_0}-2\nu_{ia}k_0)z]\Theta(z)}
\end{align}
Then  the Bourret
approximation  \eqref{Baverage} results in the following  closed expression for $\langle B \rangle$:
\begin{eqnarray}\label{Baveraged}
R_{BB}^{-1}\langle B({\bf r},z,t)\rangle
=-(k_0^2/2)(n_e/n_c)\int\int\int d{\bf r}'dz'dt'
\nonumber \\
\times R_{\sigma\sigma}({\bf r}-{\bf r}',z-z',t-t') C^*({\bf r}-{\bf r}',z-z',t-t')
\nonumber \\
\times\langle B({\bf r}',z',t') \rangle,
\end{eqnarray}
where the kernel of the response function $ R_{\sigma\sigma}({\bf x},z,t)$ is the inverse  inverse Fourier transform of \eqref{Rsigmasigma0} and  $C$ is given by \eqref{cdef}.

 We look for solution of \eqref{sigmastaraveraged} in exponential form
$\langle B\rangle \propto e^{i(\kappa z+{\bf k}\cdot {\bf r}-\omega t)}$, then the exponential time dependence
of (\ref{cdef}) allows to carry all integrations in \eqref{Rsigmasigma0} and \eqref{Baveraged} explicitly to arrive at the following
dispersion relation in dimensionless units
\begin{eqnarray}\label{dispk0contTcB}
&\Delta_B(\omega,\kappa)=
i\omega\frac{c_s}{c}+i\kappa \nonumber \\
&\qquad\qquad\qquad\qquad\qquad+
i\frac{\mu}{4}\tilde I\frac{1}{\kappa-\omega-i\mu-i\frac{1}{\tilde T_c}}%\nonumber \\
=0.
\end{eqnarray}
Here we neglected the contribution to $\kappa_B\equiv Im(\kappa)$ from diffraction %which gives negligible correction
and used the condition $c_s/c\ll 1.$
Equation (\ref{dispk0contTcB}) does not have a convective threshold (provided we neglect here light wave damping) while $\kappa_B$
has near-linear dependence on   $\tilde T_c:$ $\kappa_B\simeq \mu\tilde I\tilde T_c/4$ for $\tilde T_c< 1/\mu$ which is typical for RPA results. It suggests that we refer $\kappa_B$ as   the RPA-like branch of
instability.

Solving the  equations (\ref{dispk0contTc}) and (\ref{dispk0contTcB}) numerically for $\kappa$ allows to find   $\kappa_\sigma$ and $\kappa_B$, respectively, for given $\omega$.
We choose $\omega=0.5$ in  (\ref{dispk0contTc}) and $\omega=0$ in (\ref{dispk0contTcB})  to maximize $\kappa_\sigma$ and $\kappa_B$, respectively.
 %For  $\tilde T_c\sim 1$
Figures \ref{fig:figkappa}a and \ref{fig:figkappa}b show  that the analytical expression $\kappa_B+\kappa_\sigma$ is a reasonably good approximation for numerical value of $\kappa_i$ above the convective threshold (\ref{I0convthresh})
for $\tilde T_c\lesssim 0.1$ which is the main result of this paper. Below this threshold  analytical and numerical results are only in qualitative agreement and we replace
$\kappa_B+\kappa_\sigma$ by $\kappa_B$ because $\kappa_\sigma<0$ in that case.

The qualitative explanation why $\kappa_B+\kappa_\sigma$ is a surprisingly good approximation to $\kappa_i$ is based on the following argument.
First imagine that $B$ propagates linearly and not coupled to the fluctuations of $\sigma^*$, so its
source is  $\sigma^* E\to \langle \sigma^* \rangle  E$ in r.h.s of (\ref{EBeq2}). %If  $\langle \sigma^* \rangle$ were independent of $z$, then so is  $\langle |B|^2 \rangle$.
If
$ \langle \sigma^*  \rangle \propto e^{\kappa_\sigma z}$ grows slowly with $z$ (i.e. if $ \langle \sigma^*  \rangle$ changes a little over the speckle length $L_{speckle}$ and time $T_c$),
then so will  $\langle |B|^2 \rangle$ at the rate $2\kappa_\sigma$. But if the
total linear response $R_{BB}^{tot}$ ($R_{BB}^{tot}$ is the renormalization of bare response $R_{BB}$ due to the coupling in r.h.s of  (\ref{EBeq2})) is unstable  % when coupling to $\sigma^* $
%fluctuations is restored,
then its growth
rate gets added to $\kappa_\sigma$ in the determination of $\langle |B|^2 \rangle$ since in all theories which allow
factorization of 4-point correlation function into product of 2-point correlation functions,
$\langle B(1)B^*(2)\rangle = R^{tot}_{BB} (1,1')S(1',2')R^{tot \, *}_{BB} (2',2)$. Here
$S(1,2) \equiv \langle\sigma^* (1)\sigma (2)\rangle \langle E(1)E^*(2) \rangle  \simeq \langle\sigma^*  (1)\rangle\langle\sigma (2)\rangle \langle E(1)E^*(2) \rangle$
and $``1", \, ``2"$ etc. mean %the respective values
a set of all spatial and temporal arguments.

\section{Convective instability versus absolute instability}
\label{sec:ConvectiveAbsolute}

In this Section we show that the dispersion relations (\ref{dispk0contTc}) and (\ref{dispk0contTcB})
 predict absolute instability for large intensities.
We first consider  the dispersion relation  (\ref{dispk0contTc}) which has branch cut in the complex $\kappa$-plane connecting two branch points $\kappa_1=1-\omega\frac{c_s}{c}-2i\frac{c_s}{c}\frac{1}{\tilde T_c}$ and
$\kappa_2=-\omega\frac{c_s}{c}-2i\frac{c_s}{c}\frac{1}{\tilde T_c}$.

Absolute  instability occurs if the contour $Im(\omega)=const$ in the complex $\omega$-plane cannot be moved down to real $\omega$ axis because of pinching of two solutions of  (\ref{dispk0contTc}) in the complex $\kappa$-plane
\cite{Briggs1964},\cite{PitaevskiiLifshitzPhysicalKineticsBook1981}. To describe instability one of these solutions must cross the
real axis in $\kappa$-plane as the contour  $Im(\omega)=const$ is moving down. The pinch occurs provided
\begin{align} \label{deltasigmapinch}
\frac{\partial \Delta_\sigma(\omega,\kappa)}{\partial \kappa}=0.
\end{align}

The pinch condition \eqref{deltasigmapinch} together with the requirement of crossing the
real axis in $\kappa$-plane result in
\begin{equation}\label{pinchconditionrho}
\kappa=\frac{1}{2}+\frac{1}{2} i \sqrt{\mu \tilde I-1 }-\frac{c_s}{c}  {\omega }-\frac{c_s}{c}  {\frac{2i}{\tilde T_c} }.
\end{equation}
Taking \eqref{pinchconditionrho} together with
 $\Delta_\sigma(\omega,\kappa)=0$ from   (\ref{dispk0contTc}) at the absolute instability threshold $Im(\omega)=0$  gives the transcendental  expression
\begin{align} \label{deltaabsthresh}
&\mu-\frac{1}{2}(\mu\tilde I_{absthresh} -1)^{1/2}+\frac{c_s}{c}  {\frac{2}{\tilde T_c} }\nonumber \\
&-\frac{1}{2}  \mu\tilde  I_{absthresh}  \arctan{}\left[(\mu\tilde I_{absthresh} -1)^{-1/2}\right]=0
\end{align}
for the  absolute instability threshold intensity $\tilde I_{absthresh}$.
Assuming  $\mu\tilde I_{absthresh}\gg 1$ we obtain from \eqref{deltaabsthresh} the explicit expression for the CBSBS absolute instability threshold%
\begin{eqnarray}\label{I0absthresh}
\tilde I_{absthresh}= \mu+3\mu^{-1}+O(\mu^{-3})+O(\tilde T_c^{-1} c_s/c).
\end{eqnarray}

The absolute instability threshold for the second RPA-like branch \eqref{dispk0contTcB} is obtained similarly  with the pinch condition
%
%\begin{align} \label{deltasigmapinch}
$\frac{\partial \Delta_B(\omega,\kappa)}{\partial \kappa}=0.
$
%\end{align}
It gives the absolute instability threshold for RPA-like branch of instability
 \begin{eqnarray}\label{I0absthreshB}
\tilde I_{absthresh,B}= \mu\left (1+\frac{1}{\mu\tilde T_c}\right )^2.
\end{eqnarray}
For $\tilde T_c\lesssim 1$, the threshold \eqref{I0absthresh} is lower than \eqref{I0absthreshB} thus \eqref{I0absthreshB} can be ignored.

For $\mu\gg 1  $ the absolute threshold (\ref{I0absthresh}) reduces to the coherent absolute BSBS instability threshold
\begin{eqnarray}\label{I0absthreshcoherent}
\tilde I_{absthreshcoherent}= \mu.
\end{eqnarray}

For typical experimental condition $\mu \gtrsim 5$. Then the absolute instability threshold  \eqref{I0absthresh} is significantly above the convective instability threshold \eqref{I0convthresh}.  Thus in simulations described below we emphasize the convective regime and assume $\tilde I$ to be below the absolute threshold.

\section{Numerical simulations}
\label{sec:Numericalsimulations}

We performed two types of simulations.
First type is $3+1D$ simulations (three spatial coordinates $\bf r$, $z$ and $t$)  of Eqs. (\ref{EBeq2}), (\ref{sigma1}) and \eqref{Elin} with the boundary and initial conditions (\ref{phik}),\eqref{phiktophat}
in the limit $c\to \infty$
 (i.e., setting $c^{-1}=0$ in (\ref{EBeq2}) and (\ref{sigma1})). It implies that the phases $\phi_{{\bf k}_j}(t-z/c)$ in \eqref{Elin} become only $t$-dependent,
$\phi_{{\bf k}_j}(t)$. That formal limit $c\to \infty$, is consistent provided $c T_c \gg L_{speckle}$.
Then in the linear instability regime, the laser field, $E$, at any time may be obtained by propagation from $z = 0$ while the scattered light field, $B$ is obtained by backward propagation from $z = L_z$.
Time scales are now set by the minimum of $T_c$ and the acoustic time scale for the density $\sigma^*$. We performed numerical simulation of $B$ and $\sigma^*$ via a split-step (operator splitting) method. $E$ advances only due to
diffraction and is determined exactly by \eqref{Elin}. For given $\sigma^*$, $B$ is first advanced due to diffraction in transverse Fourier space, and then the source term (r.h.s. of (\ref{EBeq2}) which is $\propto \sigma^* E$) is added for all ${\bf r}=(x,y)$. The density $\sigma^*$ is evolved in
the strong damping approximation in which the $d/dz$ term is omitted from equation \eqref{sigma1}. In the regimes of interest, in particular  near the collective threshold \eqref{I0convthresh} regime, the dimensionless
damping coefficient in (\ref{sigma1}) increases with acoustic Landau damping coefficient, and even for its physically smallest value of $0.01$, the scaled damping $\mu$ is approximately $5$ while $d/dz$ is either $\simeq\kappa_i$ or
$1/10$ (an inverse speckle length). So given $E$ and $B$, $\sigma^*$ may be advanced in time at each $z$, for each transverse Fourier mode, or since the transverse Laplacian term is estimated as
unity in magnitude (base on the speckle width estimate of   $F\lambda_0$), $\sigma^*$ may be approximately advanced at each spatial lattice point.

Second type is $2+1D$  simulations  (two spatial coordinates $x$, $z$ and $t$) of Eqs. (\ref{EBeq2}),(\ref{sigma1})  and\eqref{Elin}
with finite value $c_s/c=1/500$ (the typical value for the experiment)
and modified top-hat boundary condition
\begin{eqnarray}\label{phiktophatmodified}
 |E_{\bf k}|=k^{1/2} \, const, \ k<k_m; \  E_{\bf k}=0, \ k>k_m,
\end{eqnarray}
chosen to mimic the extra factor $k$ in the
  integral over transverse direction of the full  $3+1D$ problem. That modified top hat choice  ensures that the linearized equations of that $2+1D$ problem give exactly the same analytical solutions (\ref{dispk0contTc}) and \eqref{dispk0contTcB} as for the full $3+1D$ problem.  We used again the split step method by integrating along the characteristics of  $\rho$ and $B$ and solving for the diffraction by Fourier transform in  the transverse coordinate $x$.

\begin{figure}[b]
\begin{center}
\includegraphics[width = 3.45 in]{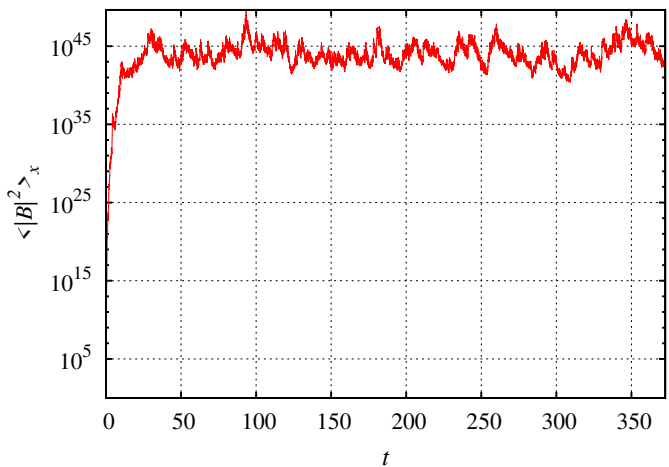}
%\vspace{0.5cm}
\end{center}
\caption{
The time dependence of  $\langle |B|^2\rangle_x$ for $3+1D$  simulation with $L_z=50$, $\mu=5.12$, $\tilde T_c=0.1$ and $\tilde I=3.0$.
$\langle |B|^2\rangle_x$ is shown at  $z=25$.
It is seen that after the initial growth, $\langle |B|^2\rangle_x$ settles at the statistical steady state with the large fluctuations around it.
The boundary condition at $z=L_z$ is $\langle |B|^2\rangle_x\sim 1$. }
\label{fig:B2timedependence}
\end{figure}

We run  simulations in the box $0<z<L_z$. For both types of simulations the boundary conditions for $B$  were set at  $z=L_z$. We take these boundary conditions as the  Fourier modes of $B({\bf r},z=L_z,t)$ in $\bf r$
with random time-independent phases. These modes correspond to the random seed from the thermal fluctuations. The boundary conditions for $\rho$ were set to be zero. As the time progress from the beginning of each simulation, both $|\rho|$ and $|B|^2$ grow
until reaching the statistical steady state if the $\tilde I$ is below the threshold of absolute instability \eqref{I0absthresh}. Figure \ref{fig:B2timedependence} shows a typical time dependence of $\langle |B|^2\rangle_x$, where $\langle \ldots\rangle_x$ means averaging over the transverse coordinate $x$.
Because we solve linear equations (\ref{EBeq2}) and (\ref{sigma1}), the maximum value of $\langle |B|^2\rangle_x$ grows  if we increase $L_z$ as well  as the boundary condition $B({\bf r},z=L_z,t)$ is defined up to the multiplication by the arbitrary constant.  $z$-dependence of $\langle |B|^2\rangle_x$ in the statistical steady state follows the exponential law $\langle |B|^2\rangle_x\propto e^{-2\kappa_iz}$ well inside the interval  $0<z<L_z$. Near the boundaries $z=0$ and $z=L_z$ there are short transition layers before solution settles at $e^{-2\kappa_iz}$ law inside that interval. The particular form of the boundary conditions for $B$ and $\rho$ affect only these transition layers while  $e^{-2\kappa_i z}$ law is insensitive to  them.

To recover $\kappa_i$ with high precision we performed simulations for long time after reaching statistical steady state and average $\langle |B|^2\rangle_x$ over that time at each $z$ (i.e. we assumed ergodicity). E.g. for $\tilde T_c=0.1$ (the time the laser light
travels along $\simeq 5$ laser speckles) we use 256 transverse Fourier modes and  discrete steps $\Delta z = 0.15$ in dimensionless units %(see definition of these units below)
with the
typical
length of the system  $L_z = 50$ ($\simeq 5$ speckle lengths)  and a time step $\Delta t = \Delta z c_s/c$. For this particular set of parameters it implies $\Delta t=3\cdot 10^{-4}$.
For  simulation we typically wait $\sim 10^5-10^6$ time steps  to achieve a robust statistical steady state and then average over
another $\sim 10^5-10^6$ time steps (together with averaging over the transverse coordinates)
%$\sim 10^6$ time steps
to find $\kappa_i$ with high precision.
Figures \ref{fig:figkappa}a and \ref{fig:figkappa}b show $\kappa_i$ extracted from  $3+1D$ and $2+1D$ simulations, respectively.

For the practical purposes it is also interesting to estimate the time $t_{ini}$ at which the initial thermal fluctuations of $|B|^2$ are amplified by $\sim e^{20}$
to reach the comparable intensity with the laser pump. We obtained from simulations that $\tilde t_{ini}\sim 0.7$ for $L_z\simeq$ two laser speckles   (relevant for gold plasma in ICF experiments and
corresponds to $L_z \simeq 22$ in dimensionless units),
 $\tilde I=3$ and $\tilde T_c=0.1$. In dimensional units for NIF conditions $t_{ini}\sim 20$ps which is well below hydrodynamic time (several hundreds of ps).

Figure \ref{fig:figcorrelation} shows normalized autocorrelation functions $\langle E({\bf r},z,t)E^*({\bf r},z,t+\Delta t)\rangle $,
$\langle B({\bf r},z,t)B^*({\bf r},z,t+\Delta t)\rangle $ and $\langle \sigma({\bf r},z,t)\sigma^*({\bf r},z,t+\Delta t)\rangle $
%in time
for $\tilde T_c=0.1$. It is seen that
the correlation times for $E$ and $B$ are similar
%close to each other
while the correlation time for $\sigma$ is much larger, the more so the smaller $\tilde T_c$. This justify the use the analytical approximations of the Section \ref{sec:Linearinstability}.

\section{Applicability of the dispersion relation and estimates for  experiment}
\label{sec:Applicability}

The applicability conditions of the Bourret approximation used in derivation of (\ref{dispk0contTc}) and (\ref{dispk0contTcB}) in the dimensionless units are
\begin{equation}\label{deltaomegadeltaomega}
\Delta \omega_B\Delta \omega_\sigma \gg\gamma_0^2.
\end{equation}
and $\Delta \omega_B\gg (c/c_s)|\kappa_B|$ as well as $\Delta \omega_\sigma\gg\mu$.
Here $\gamma_0$ is the temporal growth rate of the spatially homogeneous solution given by $\gamma_0^2=(1/4)(c/c_s)\mu\tilde I.$  Also
$\Delta \omega_\sigma=1/\tilde T_c$ is the bandwidth for $\sigma$  and  $\Delta \omega_B$ is the effective bandwidth for $B$.  $\Delta \omega_B$ is dominated by the diffraction in (\ref{EBeq2})
giving in the dimensionless units $\Delta \omega_B=c/c_s$.   Then (\ref{deltaomegadeltaomega}) reduces to  $\tilde T_c\ll 4/(\mu \tilde I)$ and $|\kappa_B|\ll 1$.
Together with the condition $T_c\gg L_{speckle}/c$ used in the derivation of (\ref{dispk0contTc}) and assuming that $\tilde I\simeq \tilde I_{convthresh}$, it gives a double inequality $(7\pi/2)(c_s/c)\ll \tilde T_c\ll \pi/\mu$
which can be well satisfied for $\mu\simeq 5$, i.e. for $\nu_{ia}\simeq 0.01$ as in
gold ICF plasma but not for $\mu\simeq 50$ as in low ionization number $Z$ ICF plasma. Also $|\kappa_B|<1$ implies that $\tilde I >\tilde I_{convthresh}$  because otherwise,
below that threshold, $\kappa_B\sim -\mu$ which would
contradict $|\kappa_B|<1$.
All these conditions are  satisfied for $\tilde T_c\lesssim 1/4$ for the parameters of Figure \ref{fig:figkappa} with $\tilde I=2$ or $\tilde I=3$ (solid lines in Figure \ref{fig:figkappa}) but not for $\tilde I=1$
(dashed lines in Figure \ref{fig:figkappa}). Additionally, an estimate for $T_c\ll t_{sat}$ from the linear part of the theory of Ref. \cite{MounaixPRL2000} results in the condition
$\tilde T_c\ll 8\tilde I/\mu$ which is less restrictive than above conditions.
  These estimates are consistent with the observed agreement between  $\kappa_i=\kappa_\sigma+\kappa_B$ and $\kappa_i$ from simulations
(filled circles in Figures \ref{fig:figkappa}) for $\tilde I$ above the threshold (\ref{I0convthresh}).
We conclude from Figures \ref{fig:figkappa} that the applicability condition for  the Bourret approximation is close to the domain of $\tilde T_c$ values for which $\kappa_i=\kappa_\sigma+\kappa_B$.

\begin{figure}[b]
\begin{center}
\includegraphics[width = 2.95 in]{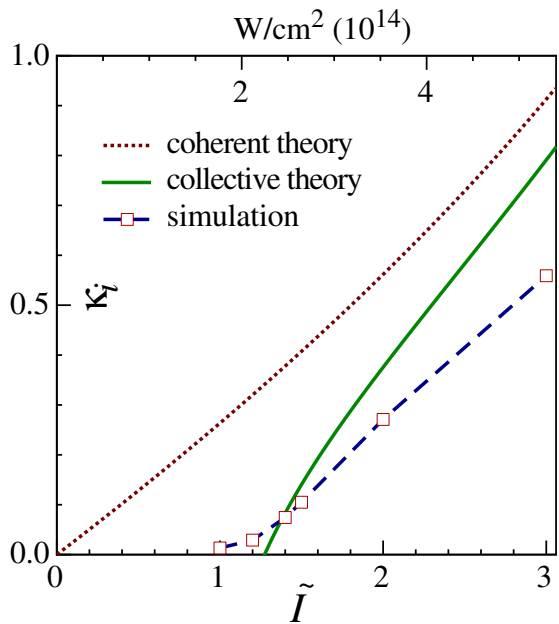}
%\vspace{0.5cm}
\end{center}
\caption{
$\kappa_i$ vs. $\tilde I$ for $\mu=5.12$  obtained from $3+1D$ simulations  (squares connected by dashed line,  $c_s/c=0$ and limit $\tilde T_c\to 0$ taken by extrapolation from $\tilde T_c\ll 1$),
analytical result $\kappa_\sigma$ (solid curve) and coherent laser beam increment  $\kappa_{coherent}$ (dotted curve). The scaled dimensionless laser intensity $\tilde I$
and damping rate  $\mu$  in units of acoustic propagation time along a speckle are defined in (\ref{Itildedef}).
Upper grid corresponds to laser intensity in dimensional units $I_{physical}\propto T_e/ \lambda_0^2$ for  NIF parameters and gold plasma
$T_e\simeq 5\mbox{keV}, \ F=8,\  \ n_e/n_c=0.1, \nu_{ia}=0.01, \  \lambda_0=351 \mbox{nm}$.
}
%\label{fig:figcorrelation}
\label{fig:kappaI}
\end{figure}

 For
nominal NIF parameters  \cite{Lindl2004,LushnikovRosePlasmPhysContrFusion2006}, $ F=8,\  \ n_e/n_c=0.1, \ \lambda_0=351 \mbox{nm}$, $c_s=6\times 10^{7}\ \mbox{cm s}^{-1}$ and
electron plasma temperature $T_e\simeq 2.6\mbox{keV}$ ($T_e$ was recently updated from the old standard value $T_e\simeq 5\mbox{keV}$ \cite{RosenInvitedAPRPlasma2011}),
we obtain from (\ref{I0convthresh}) that
$I_{convthresh}\simeq 1.1\times 10^{14}\mbox{W}/\mbox{cm}^2$ for gold plasma with   $\nu_{ia}\simeq 0.01$ which is in the range of NIF  single polarization intensities.
Fig. \ref{fig:kappaI} shows $\kappa_i$ in the limit $c_s/c=0, \ \tilde T_c\to 0$ from simulations,
analytical result $\kappa_\sigma$ ($\kappa_B=0$ in that limit) and the instability increment of the coherent laser beam $\kappa_{coherent}=\mu/2-(\mu^2-\mu\tilde I)^{1/2}/2$ (see e.g. \cite{Kruer1990}).
It is seen that the coherent increment significantly overestimates numerical increment especially around $I_{convthresh}.$ The convective increment
$\kappa_i$ has a significant dependence on $\tilde T_c$ if we include the effect of finite $c/c_s=500$ and finite $\tilde T_c$ as in Fig. \ref{fig:figkappa}b.
Current
NIF 3{\AA} beam smoothing design has $T_c\simeq 4$ps implying $\tilde T\simeq 0.15$. In that case Fig. \ref{fig:figkappa}b
shows  that there is a significant (about 5 fold) change in $\kappa_i$ between $\tilde I=1$ and $\tilde I=3$.
Similar estimate for KrF lasers ($\lambda_0=248 \mbox{nm}, \ F=8, \ T_c=0.7$ps) gives $\tilde T_c =0.04$  which results in a significant ($40\%$) reduction of $\kappa_i$ for $\tilde I=3$
compare with above NIF estimate.

The BSBS threshold may be reduced  by self-induced temporal incoherence   (see e.g. \cite{SchmittAfeyan1998}), which in its linear regime, includes
collective FSBS (CFSBS) which reduces $T_c$ and laser correlation lengths. For
low $Z$ plasma, the CBSBS and CFSBS thresholds are close while the latter may be lowered by adding higher $Z$ dopant.

\section{Conclusion}
\label{sec:Conclusion}

In conclusion, we identified the collective threshold (\ref{I0convthresh}) of  BSBS instability of partially incoherent laser beam for ICF relevant plasma. Above that threshold the CBSBS increment $\kappa_i$ is well
approximated by the sum of the collective-like increment $\kappa_\sigma$
and RPA-like increment $\kappa_B$. That result is in agreement with the direct stochastic simulations of BSBS equations.
  Values of $\kappa_\sigma$ and  $\kappa_B$ are comparable above threshold while in a small neighborhood of  threshold  the value of $\kappa_i$ changes quickly
with changing either correlation time or laser intensity to pass through collective threshold. With further increase of laser intensity the absolute instability also develops above the threshold \eqref{I0absthresh}.

%%%%%%%%%%%%%%%%%%%%%%%%%%%%%%%%%%%%%%%%%%%%%%%%%%
%%%%%%%%%%%%%%%%%%%%%%%%%%%%%%%%%%%%%%%%%%%%%%%%%%
\begin{acknowledgments}
We acknowledge helpful discussions with R. Berger and N. Meezan.
P.\,L. and H.\,R. were supported by the New Mexico Consortium and Department of Energy
Award No. DE-SCOO02238 as well as by the National Science Foundation
under Grants No. PHY 1004118, and
No. PHY 1004110.
A.\,K. was partially supported by
the Program ``Fundamental problems of nonlinear dynamics'' from the RAS
Presidium and ``Leading Scientific Schools of Russia'' grant
NSh-6170.2012.2.
\end{acknowledgments}


\begin{thebibliography}{}


\bibitem{Lindl2004} J.\,D.~Lindl, {\it et al.},  Phys.\,Plasmas {\bf 11}, 339
(2004).

\bibitem{MeezanEtAlPhysPlasm2010} N.\,B.~Meezan, {\it et al.}, Phys.\,Plasmas {\bf 17}, 056304 (2010).


\bibitem{FroulaPRL2007} S. H. Glenzer {\it et. al.}, Nature Phys., {\bf 3}, 716 (2007);
D. H. Froula, {\it et al.}, Phys. Rev. Lett., {\bf 98},
085001 (2007); N. B. Meezan, {\it et al.}, Phys. Plasmas,
{\bf 17}, 056304 (2010) X. Meng, {\it et al.}, Phys. Plasmas, High Power Laser Science and Engineering, {\bf 1},
94 (2012).


\bibitem{Kruer1990} W.\,L.~Kruer, {\it The physics of laser plasma interactions},
Addison-Wesley, New York (1990).

\bibitem{RosePhysPlasm1995} H.\,A.~Rose, Phys.\,Plasmas {\bf 2}, 2216 (1995).

\bibitem{GarnierPhysPlasm1999}  J.~Garnier, Phys.\,Plasmas {\bf 6}, 1601 (1999).

\bibitem{vedenov1964}
    A.\,A.~Vedenov, and L.\,I.~Rudakov, Sov.\,Phys.\,Doklady {\bf 9}, 1073 (1965); A.\,M.~Rubenchik, Radiophys. Quant. Electron. {\bf 17},
    1249 (1976); V.\,E.~Zakharov, S.\,L.~Musher, and A.\,M.~Rubenchik, Phys.\,Rep. {\bf 129}, 285 (1985).

\bibitem{PesmeBerger1994} D.~Pesme,  {\it et al.}, Natl. Tech. Inform. Document No.\,PB92-100312 (1987);         arXiv:0710.2195 (2007).


\bibitem{LushnikovRosePRL2004} P.\,M.~Lushnikov and H.\,A.~Rose, Phys.\,Rev.\,Lett. {\bf 92},
255003 (2004).

\bibitem{LushnikovRosePlasmPhysContrFusion2006}
P.\,M.~Lushnikov and H.\,A.~Rose,
        Plasma\,Phys. Controlled Fusion {\bf 48}, 1501 (2006).


%\bibitem{Weaver2007} J. L. Weaver {\it et. al.}, Phys. Plasm. {\bf 14},
%056316 (2007).



\bibitem{RoseDuBois1994}H.\,A.~Rose and D.\,F.~DuBois, Phys.\,Rev.\,Lett. {\bf 72}, 2883 (1994).

\bibitem{LushnikovRoseArxiv2007}
P.\,M.~Lushnikov and H.\,A.~Rose, arXiv:0710.0634  (2007).




\bibitem{NiemannPRL2008} C.~Niemann, {\it et al.},  Phys.\,Rev.\,Lett. {\bf 100}, 045002 (2008).

\bibitem{RemarkReplete} The literature is replete with multi-dimensional simulations of SBS,  with models which are similar to the one used in our work
 (see. e.g. \cite{SchmittAfeyan1998,MassonLabordePesmeEtAlJDePhys2006,PesmeEtAlPRL2000}).
  However,
  the other works  emphasize nonlinear regimes with competing instabilities, such as BSBS and filamentation, while we apply our theory and simulation to strictly linear BSBS regime.



\bibitem{RoseMounaixPhysPlasm2011}H.\,A.~Rose and Ph.~Mounaix, Phys.\,Plasmas {\bf 18}, 042109  (2011).

\bibitem{Briggs1964} A. Bers, pp. 451-517,
In {\it Handbook of plasma physics}, Eds. M.N Rosenbluth, {\it at. al.}, North-Holland (1983).

\bibitem{PitaevskiiLifshitzPhysicalKineticsBook1981}  L. P. Pitaevskii, and E.M. Lifshitz,
{\it Physical Kinetics: Volume 10},
Butterworth-Heinemann, Oxford (1981).


\bibitem{MounaixPRL2000} Ph.~Mounaix, {\it et al.}, Phys.\,Rev.\,Lett. {\bf 85},
4526 (2000).

%\bibitem{BergerPrivate2007} R.L. Berger, Private communication (2007).



\bibitem{DuBoisBezzeridesRose1992} D.\,F.~DuBois, B.~Bezzerides, and H.\,A.~Rose,  Phys.\,of Fluids B: Plasma Physics {\bf 4}, 241 (1992).




\bibitem{LehmbergObenschain1983}R.\,H.~Lehmberg and S.\,P.~Obenschain, Opt.\,Commun. {\bf 46}, 27 (1983).

\bibitem{polarization} Subsequent analysis can be easily generalized to include polarization smoothing \cite{Lindl2004}.



\bibitem{RoseDuBois1993} H.\,A.~Rose and D.\,F.~DuBois, Phys.\,of Fluids
   {\bf B5}, 3337 (1993).


\bibitem{RosenInvitedAPRPlasma2011} %M.D Rosen, Invited talk,  53rd Annual Meeting of the APS Division of Plasma Physics,
%Salt Lake City  (2011).
M.D. Rosen,{\it et al.}, High Energy Density Physics {\bf 7},  180 (2011).


\bibitem{absolutecomment} This expression is valid for $\mu>\pi/4$, while for $\mu\le\pi/4$ the convective threshold coinsides with the absolute instability threshold.


\bibitem{SchmittAfeyan1998}A.\,J.~Schmitt and B.\,B.~Afeyan, Phys.\,Plasmas {\bf 5}, 503 (1998).

\bibitem{MassonLabordePesmeEtAlJDePhys2006}
P.E. Masson-Laborde, {\it et al}.,  J. De Physique IV {\bf 133}, 247 (2006).

\bibitem{PesmeEtAlPRL2000} D.~Pesme, {\it et al.},   Phys.\,Rev.\,Lett. {\bf 84}, 278 (2000); A.\,V.~Maximov, {\it et al.}, Phys.\,Plasmas {\bf 8}, 1319 (2001);
P.~Loiseau, {\it et al.}, Phys.\,Rev.\,Lett. {\bf 97}, 205001 (2006).









%\end{references}
\end{thebibliography}
\end{document}